\documentclass{epl}

\usepackage{amsmath,amssymb}
\usepackage{graphics,epsfig}

\title{Multi-terminal transport through a quantum dot in the Coulomb blockade regime}
\shorttitle{Multi-terminal transport through a quantum dot}
\author{R. Leturcq\inst{1}\thanks{E-mail: \email{leturcq@phys.ethz.ch}} \and D. Graf\inst{1} \and T. Ihn\inst{1} \and K. Ensslin\inst{1} \and D. D. Driscoll\inst{2} \and A. C. Gossard\inst{2}}
\shortauthor{R. Leturcq \etal}
\institute{
  \inst{1} Solid State Physics Laboratory - ETH Z\"urich, 8093 Z\"urich,  Switzerland\\
  \inst{2} Materials Department, University of California - Santa Barbara, Ca 93106, USA}

\begin{document}

\maketitle

\begin{abstract}
Three terminal tunnelling experiments on quantum dots in the Coulomb-blockade regime allow a quantitative determination of the coupling strength of individual quantum states to the leads. Exploiting this insight we have observed independent fluctuations of the coupling strengths as a function of electron number and magnetic field due to changes in the shape of the wave function in the dot. Such a detailed understanding and control of the dot-lead coupling can be extended to more complex systems such as coupled dots, and is essential for building functional quantum electronic systems.
\end{abstract}

In a standard two-terminal experiment with a single quantum dot in the Coulomb blockade regime \cite{Kouw01}, the current in a conductance resonance is determined by the average coupling of the electron wave function in the dot with the corresponding wave functions in both leads. In the linear regime, such an experiment does not allow to determine the individual coupling of the wave function in the dot to each terminal. Here we demonstrate that in the single-level tunnelling regime of the Coulomb blockade it is possible to deduce the individual coupling strengths from the dot to the leads if three or more terminals are connected to the dot. It is possible to determine the conductance matrix of the quantum dot, and to calculate the individual tunnelling rates from the dot to each lead. For weak coupling, the magnitude of the tunnelling rates of a given terminal is found to vary independently of the two other tunnelling rates when the number of electrons in the dot is changed. This result can be related to the chaotic nature of the wave function in the dot. The fluctuations of the shape of the wave function in the dot due to quantum interference is directly observed via the magnetic field dependence of the coupling strengths. Finally, level broadening beyond thermal effects becomes visible as the coupling strength of a single lead increases. This set-up allows then to tune the tunnelling rates from the dot into the three terminals individually.

\begin{figure}
\onefigure[width=14cm]{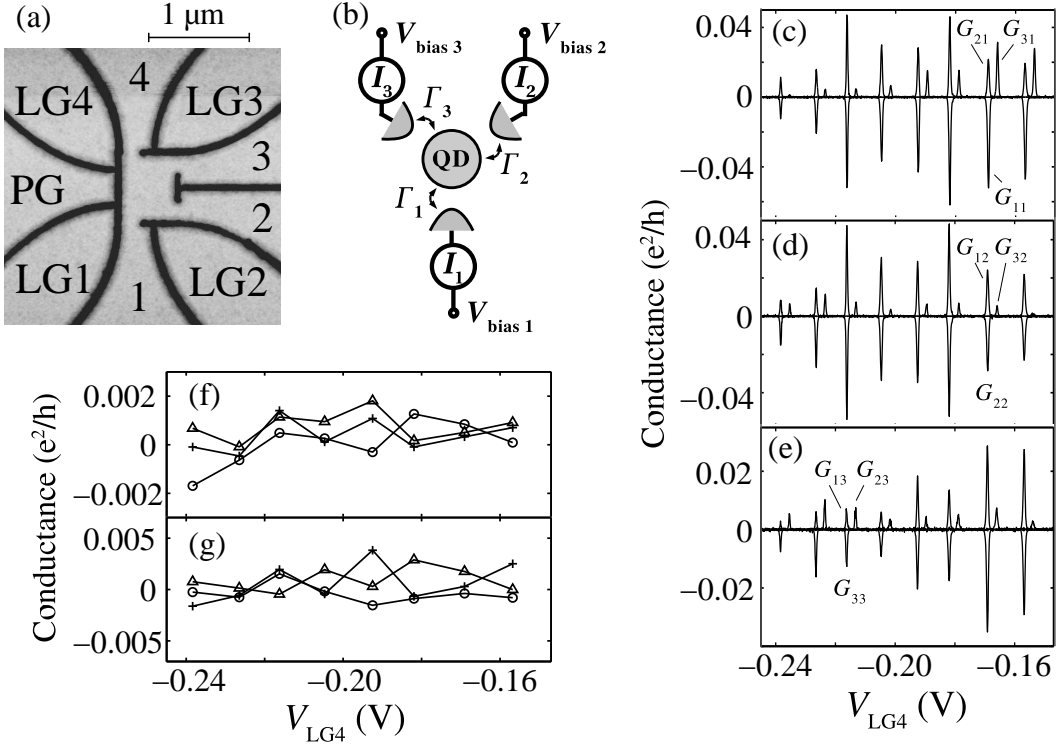}
\caption{a) Micrograph of the four-terminal quantum dot, the black lines being the oxide lines. The four leads (labelled 1 to 4) to the dot can be tuned through the four lateral gates LG1 to LG4. The plunger gate PG tunes the number of electrons in the dot. b) Measurement set-up using three of the four terminals. The quantum dot (QD) is connected to the three leads through tunnel barriers with tunnelling rates $\Gamma_1$ to $\Gamma_3$. c)-e) Measurement of Coulomb blockade resonances in the three configurations, when applying a bias voltage to one lead, the others being grounded. For each plot, the negative conductance corresponds to the current in the biased lead (1 in c), 2 in d) and 3 in e)), and the positive conductances to the two currents in the grounded leads. One of them is laterally offset by a constant of +3 mV to make the presentation more transparent. f) Current sum rule for the conductance at the resonances presented in panels c) to e), obtained by adding the terms of each column of the conductance matrix: $G_{11}+G_{21}+G_{31}$ (+), $G_{12}+G_{22}+G_{32}$ (o) and $G_{13}+G_{23}+G_{33}$ ($\Delta$). g) Voltage sum rule obtained by adding the terms of each row of the conductance matrix: $G_{11}+G_{12}+G_{13}$ (+), $G_{21}+G_{22}+G_{23}$ (o) and $G_{31}+G_{32}+G_{33}$ ($\Delta$).}
\label{fig1}
\end{figure}

The sample has been fabricated on an AlGaAs-GaAs heterostructure containing a two-dimensional electron gas (2DEG) 34 nm below the sample surface. A back gate situated 1.4 $\mu$m below the 2DEG allows to tune the electron density. All measurements presented here were performed at a back gate voltage of -1.4 V, giving a 2DEG density of $3.7 \times 10^{11}$ cm$^{-2}$ and a mobility of 200'000 cm$^2$/Vs at $T=4.2$ K. The surface of the heterostructure was locally oxidised by applying a voltage between the conductive tip of an atomic force microscope (AFM) and the 2DEG \cite{Held01}. The electron gas is depleted below the oxide lines. This patterning technique was used in other studies for defining  high-quality quantum dots \cite{Lusc01,Fuhr01}. The details of the fabrication process, which are crucial for the high electronic quality of the quantum dot, are described in ref.~\cite{Fuhr04}. Figure~\ref{fig1},a) shows the oxide lines defining the quantum dot. The AFM image was taken with an unbiased tip directly after the oxidation process. The width of the four quantum point contacts connecting the dot to the four reservoirs numbered 1 through 4 is controlled by voltages applied to the lateral gate electrodes LG1 through LG4. The number of electrons in the dot can be tuned via the lateral plunger gate PG. In this paper we focus on the dot being connected to three terminals. The point contact connecting the dot to reservoir 4 is completely closed, and gate LG4 can be used for controlling the number of electrons in the dot. Measurements of Coulomb diamonds reveal a charging energy $E_C \approx 0.5$ meV and an average single-particle level spacing $\Delta \approx 35$ $\mu$eV, compatible with an electronic dot area of 400 nm $\times$ 250 nm estimated from the lithography pattern and the lateral depletion. An electronic temperature of 90 mK is deduced from the width of the Coulomb peaks, as it will be explained later. The quantum dot can be tuned into the quantum Coulomb blockade regime with the mean single-particle level spacing $ \Delta$ being much larger than the thermal energy $k_BT$ and the level broadening $h \Gamma$.

Figure~\ref{fig1},b) shows the measurement set-up. A dc bias voltage of 10~$\mu$V is applied to one terminal of the dot ({\it e.g.} $V_{bias1}$), while the two other terminals are grounded ({\it e.g.} $V_{bias2}=V_{bias3}=0$). Current-voltage converters are used to measure the currents through each terminal. We present the experimental results in the following way: for a bias voltage applied to one terminal, the currents through all three terminals are measured. In order to minimise the influence of possible offsets, measurements for positive and negative bias are averaged. Then, by applying the bias successively to each of the three terminals, we obtain nine different current measurements. In linear response theory, these nine currents correspond to the nine elements of the conductance matrix $\bold{G}$ of the three-terminal system:
\begin{equation*}
\left( \begin{array}{c} I_1 \\ I_2 \\ I_3 \end{array} \right) = \left( \begin{array}{c c c} G_{11} & G_{12} & G_{13} \\ G_{21} & G_{22} & G_{23} \\ G_{31} & G_{32} & G_{33} \end{array} \right) \left( \begin{array}{c} V_1 \\ V_2 \\ V_3 \end{array} \right) = \bold{G} \left( \begin{array}{c} V_1 \\ V_2 \\ V_3 \end{array} \right)
\end{equation*}
In a preliminary experiment on a strongly coupled dot some of the $G_{ij}$'s have been measured \cite{Kumar02}.

Figures~\ref{fig1},c)-e) show the nine conductances $G_{ij}$ as a function of the gate voltage $V_{LG4}$ that controls the number of electrons on the dot (in each panel, one curve is laterally offset by 3 mV for clarity). The positions of all corresponding Coulomb resonances agree within less than 1/10 of the peak width ({\it i.e.} less than 5~$\mu$eV), indicating that the same energy level in the dot is probed in all configurations. Current conservation implies $\sum_{i=1}^3 G_{ij} = 0$ for all $j$. Additionally, if all voltages are set to the same value no current should flow: $\sum_{j=1}^3 G_{ij} = 0$ for all $i$. Figures~\ref{fig1},f) and g) show that these two sum rules are obeyed by the experimental data with a relative accuracy better than 10\% of the highest current level.

In the case of very low temperature and weak coupling, one can use the theory for lowest order sequential tunnelling including interaction effects in the dot, generalised to the case of more than two terminals. Following Beenakker \cite{Beenakker01}, we find from the rate equation approach that, in the regime of weak coupling ($h \Gamma \ll k_BT$) and in the single-level transport regime ($k_BT \ll \Delta$), the elements of the conductance matrix $\bold{G}$ are given by
\begin{eqnarray}
G_{ij} & = & \frac{e^2}{4k_BT} \frac{\Gamma_i \Gamma_j}{\Gamma_1+\Gamma_2+\Gamma_3} \cosh^{-2} \left( \frac{\delta}{2k_BT} \right) \qquad \text{for $i \neq j$}\label{gij3term}\\
G_{ii} & = & -\frac{e^2}{4k_BT} \frac{\Gamma_i \left( \Gamma_j + \Gamma_k \right)}{\Gamma_1+\Gamma_2+\Gamma_3} \cosh^{-2} \left( \frac{\delta}{2k_BT} \right) \qquad \text{for $i \neq j$, $i \neq k$ and $j \neq k$}\label{gii3term}
\end{eqnarray}
with $\Gamma_k$ being the tunnelling rate from the dot to lead $k$ (see fig.~\ref{fig1},b)), and $\delta = e \alpha_{LG4} (V_{LG4,res}-V_{LG4})$, with $\alpha_{LG4}$ the lever arm of gate LG4, determined by the measurement of Coulomb diamonds. Each peak shown in fig.~\ref{fig1},c)-e) is fitted with eq. (\ref{gij3term}) or (\ref{gii3term}) in order to deduce its position, its maximum and its width. From the maxima of the peaks, we calculate the individual tunnelling rates $\Gamma_k$. The tunnelling rates depend on the overlap of the wave function in the dot with the wave function in lead k. This overlap depends on the shape of the wave function in the dot, and therefore on the interference pattern in the dot.

\begin{figure}
\centering{
\begin{tabular}{lr}
\includegraphics[width=7.1cm]{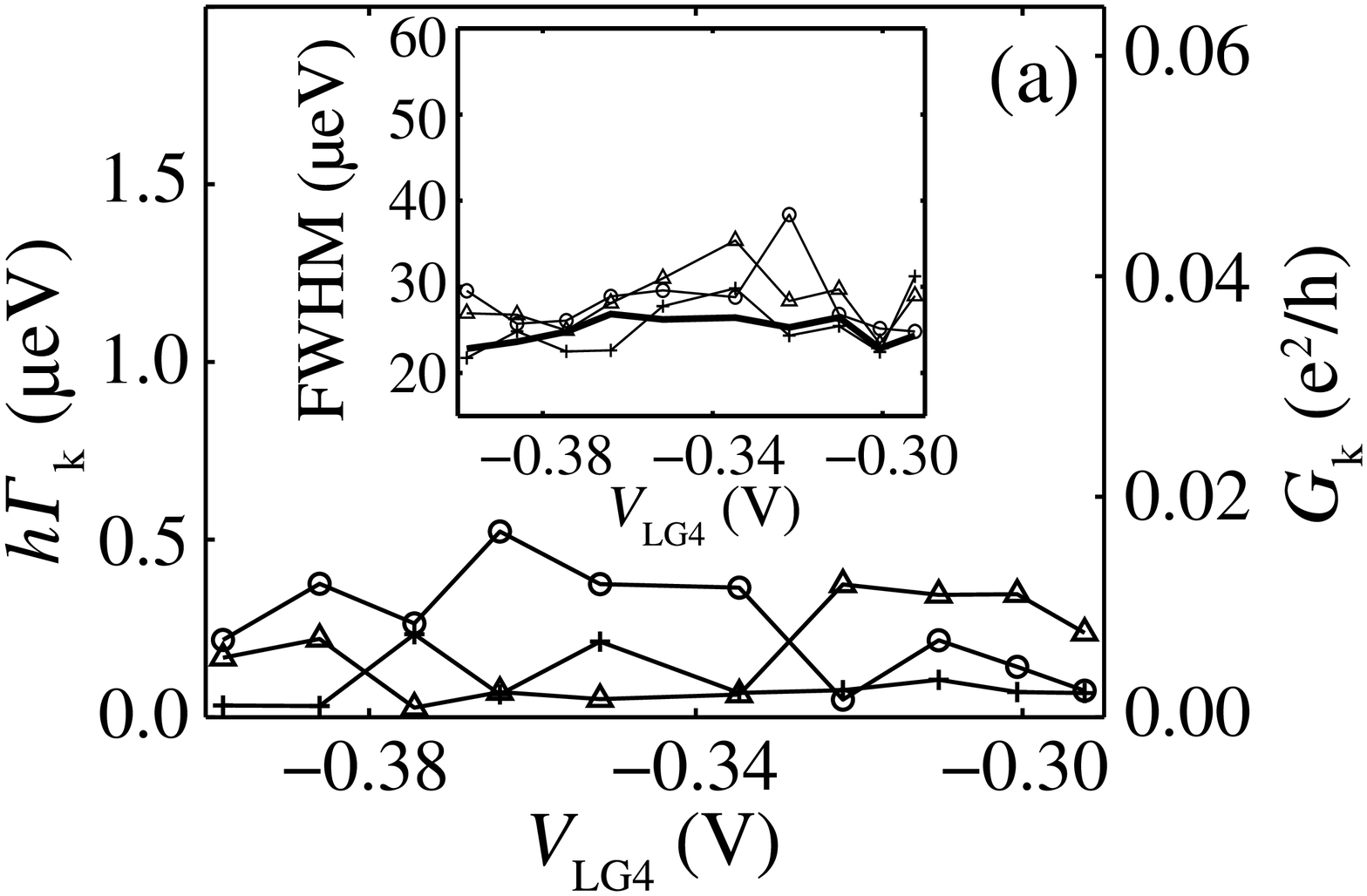}&
\includegraphics[width=6.8cm]{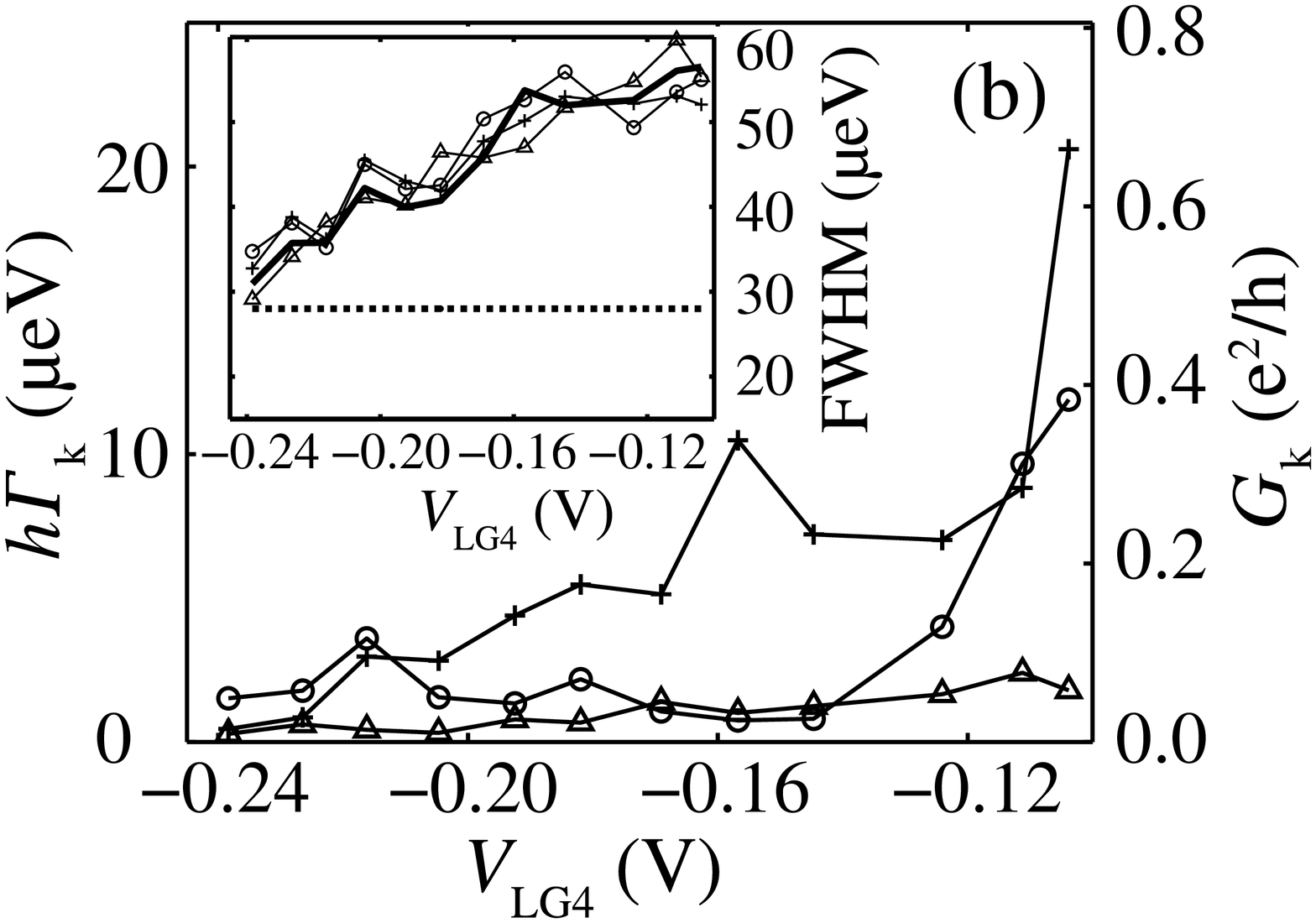}
\end{tabular}
}
\caption{Tunnelling rates $h \Gamma_k$ for individual resonances for lead 1 (+), lead 2 (o) and lead 3 ($\Delta$), and corresponding conductances $G_k$ (right hand axes). Inset: Peak width at half maximum (FWHM) for the resonance measured in $G_{11}$ (+), $G_{22}$ (o), $G_{33}$ ($\Delta$), and mean peak width of the nine conductance matrix elements (thick line). a) For low $V_{LG4}$, {\it i.e.} weak coupling. b) At higher $V_{LG4}$, {\it i.e.} intermediate coupling. Inset: The dashed thick line at 27 $\mu$eV corresponds to a thermally broadened peak at 90 mK, the temperature corresponding to the mean peak width for a more closed dot (inset of panel a)).}
\label{fig2}
\end{figure}

The tunnelling rates measured at resonance in the weak coupling regime as a function of the gate voltage $V_{LG4}$ are shown in fig.~\ref{fig2},a). Such an experimental determination of tunnelling rates is unique to a three (or more) terminal quantum dot and not possible in conventional two-terminal experiments in the linear regime. The values of the tunnelling rates  fluctuate strongly and independently \cite{Beenakkerandother}, also when the mean coupling strengths are similar for all leads (fig.~\ref{fig2},a)). Only for more positive gate voltages (fig.~\ref{fig2},b)), where the dot starts to open because of electrostatic coupling of the plunger gate to the point contacts, the tunnelling rates start to increase systematically. However, in this case, the single-level tunnelling regime might not be valid anymore because of the higher tunnelling rates, and the determination of individual tunnelling rates is no longer valid due to co-tunnelling.

\begin{figure}
\twofigures[width=7cm]{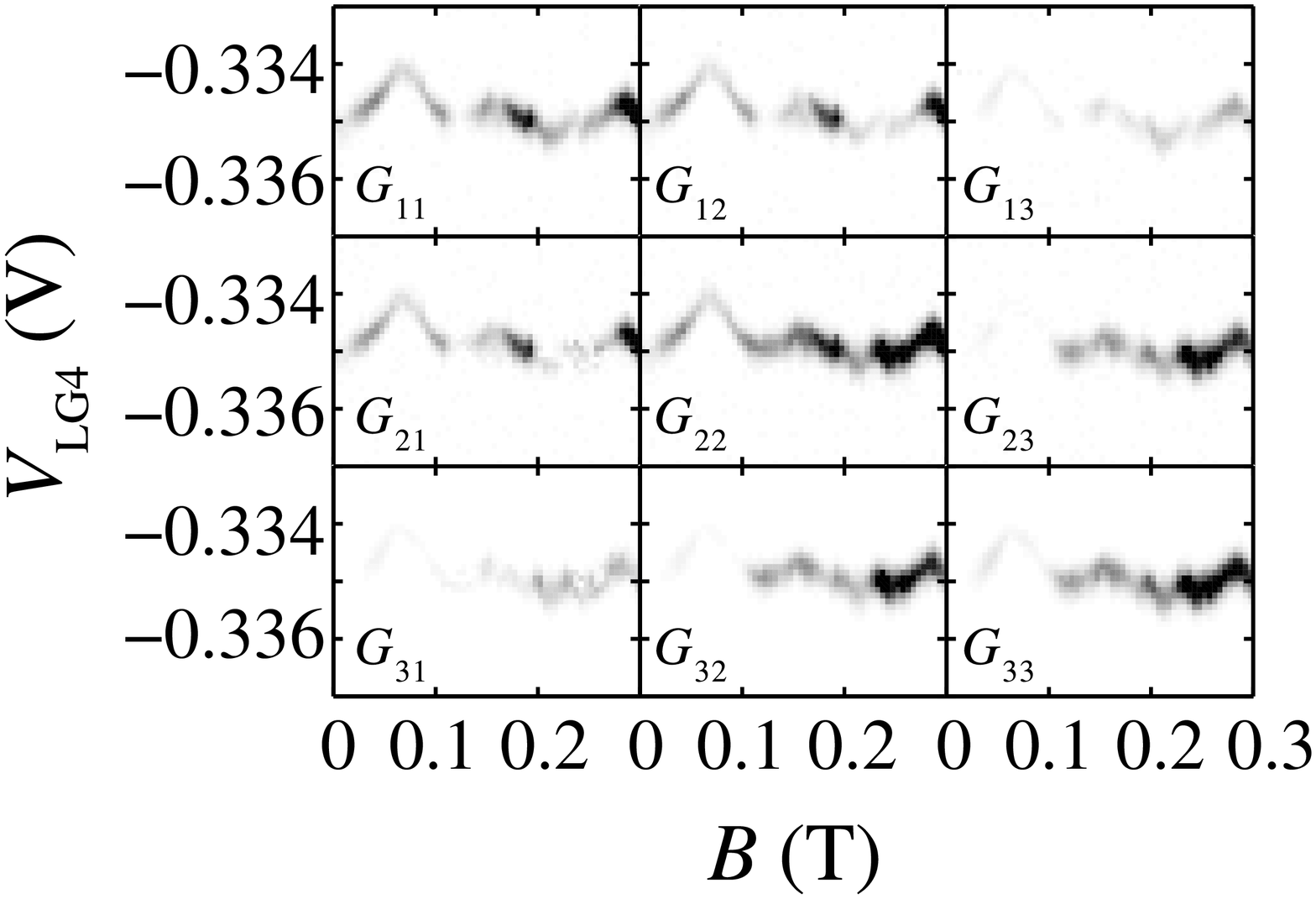}{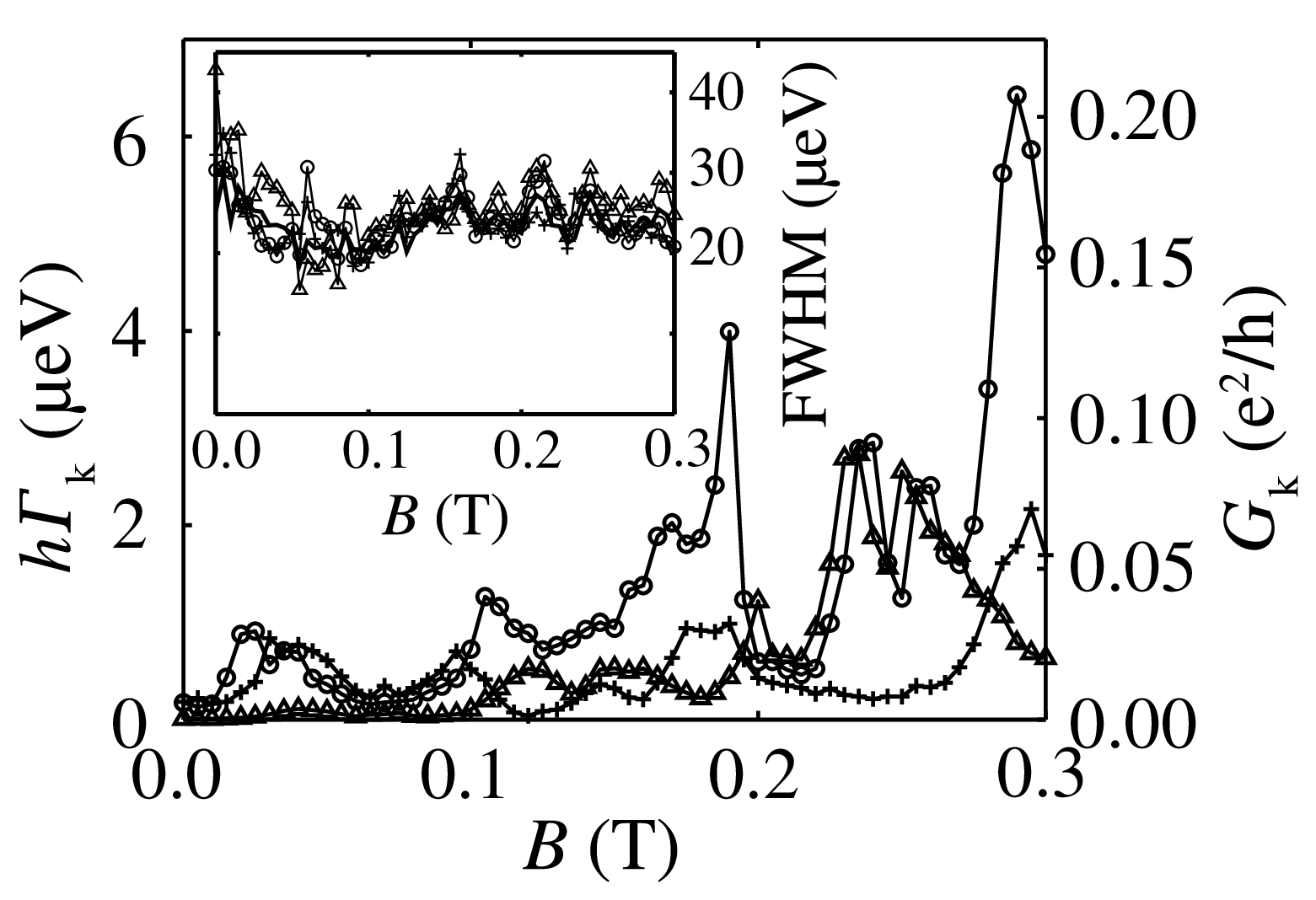}
\caption{Absolute value of the conductance matrix elements $G_{ij}$ vs. $V_{LG4}$ and magnetic field $B$ around a resonance. The conductance is represented with a linear gray scale with white corresponding to 0 and black to $\geq 0.025$ $e^2/h$.}
\label{fig3}
\caption{Tunnelling rates at a resonance $h \Gamma_k$, for lead 1 (+), lead 2 (o) and lead 3 ($\Delta$), vs. magnetic field $B$, and corresponding conductances $G_k$ (right hand axes). The magnetic field necessary to add a flux quantum through the dot area is 40 mT. Inset: Peak width at half maximum (FWHM) for the resonance measured on $G_{11}$ (+), $G_{22}$ (o), $G_{33}$ ($\Delta$), and mean of the nine conductances (thick line).}
\label{fig4}
\end{figure}

It is known that states in a dot change also as a function of magnetic field due to changes in the quantum interference pattern. We have measured the conductance matrix around one resonance as a function of the magnetic field $B$ and the gate voltage $V_{LG4}$, as shown in fig.~\ref{fig3}. The fact that the peak positions fluctuate the same way as a function of the magnetic field for all configurations confirms that the three leads couple to the same state. Following each peak, we have extracted the peak heights and plot the tunnelling rates as a function of the magnetic field in fig.~\ref{fig4}. Again, the three tunnelling rates fluctuate independently.

We like to note that, at a resonance, the $G_{ij}$ defined in eqs.~(\ref{gij3term}) and (\ref{gii3term}) are formally equivalent to the conductance matrix elements of a classical star-shaped conductance network, the three conductances connecting each lead $k$ to the center of the dot being defined as $G_k = (e^2/h) (h \Gamma_k)/(4 k_BT)$. We can therefore represent $h \Gamma_k$ in figs.~\ref{fig2} and \ref{fig4} as equivalent conductances $G_k$ (right hand axes).

Does coherent tunnelling change this analysis? It has been shown experimentally that tunnelling through a quantum dot is at least partially coherent \cite{YacobySchuster}. Coherence may have an influence on the dot conductance if interference is possible \cite{FoaTorres01}. If a particular level spacing $\delta \varepsilon$ becomes significantly smaller than the average level spacing $\Delta \approx 35$ $\mu$eV, the condition $h \Gamma > \delta \varepsilon$ leads to level mixing due to interference. For such a case our relation between the $G_{ij}$ and the $\Gamma_k$ has to be modified. However, if the coupling is weak, $h \Gamma \ll \Delta$, no interference is possible. In this case, it has been shown at least for the two-terminal experiment that the expression for coherent tunnelling reduces to the known equations for sequential tunnelling \cite{LewenkopfPC}.

Equation.~(\ref{gij3term}) implies that $G_{ij}=G_{ji}$ because $G_{ij} \propto \Gamma_i \Gamma_j$. This agrees with the generalized Onsager relations $G_{ij}(B)=G_{ji}(-B)$ at zero magnetic field. At finite magnetic field this symmetry can not be observed in general \cite{Buttiker01}, an extreme example being an electron focusing experiment \cite{vanHouten01}, and eq.~(\ref{gij3term}) may not have the most general form required for finite $B$. However, we find that in our experiment $G_{ij}(B)=G_{ji}(B)$ for magnetic fields below 0.3~T (see fig.~\ref{fig3}), which implies $G_{ij}(B)=G_{ij}(-B)$ through the Onsager relations. In the absence of a more appropriate theory we therefore employ eq.~(\ref{gij3term}) empirically also at finite $B$. Nevertheless we may ask whether the symmetry of the conductance matrix in our measurement is a general property of any three-terminal quantum dot or the result of particular microscopic properties of our system. We speculate that the observed symmetry is a general property of a multi-terminal system coupled weakly to the leads, because the tunnelling matrix elements depending on the overlap of dot and lead wave functions are even in magnetic field. In lowest order in the tunnelling the $G_{ij}$ will therefore be symmetric in $B$. However, we believe that this question has to be settled eventually with new experiments and a thorough theoretical analysis.

From the measurement of individual dot-lead coupling strengths, we can address three questions. How can we probe changes in the wave function in the dot? How is the level broadening affected by opening a single contact? And how can the coupling strengths be monitored on a quantitative level?

Peak height fluctuations attributed to fluctuations of the shape of quasi-bound states in chaotic dots \cite{Kouw01} have been extensively studied in two-terminal quantum dots \cite{ChangFolk,PatelFolk}. Calculations based on random matrix theory are in reasonable agreement with experimental results \cite{FoaTorres01,ChangFolk,Jala01}. However, two-terminal experiments can only give information on the global conductance of the entire system. In a three-terminal setup this picture can be probed more directly by looking at the spatial distribution of the wave function. In our experiment, these spatial fluctuations are observed as uncorrelated tunnelling rates because the distance between leads ($\approx 400$ nm) is much larger than the Fermi wave length ($\approx 40$ nm) \cite{LewenkopfPC}. A magnetic field perpendicular to the 2DEG changes the shape of the wave function, as it has been suggested for the interpretation of the peak height fluctuations as a function of the magnetic field in two-terminal experiments \cite{ChangFolk}. In our experiment, independent fluctuations of the individual tunnelling rates on a scale of 40 mT, corresponding to one flux quantum added within the dot area, are a direct consequence of the spatial fluctuations of the wave function due to changes of the interference pattern in the dot.

In order to check the influence of the coupling to the leads on the level broadening, an effect that is beyond eqs. (\ref{gij3term}) and (\ref{gii3term}), we have also carried out the analysis of the peak width for all configurations. For weak coupling, $h \Gamma \ll k_BT \ll \Delta \ll E_C$, the peak width is approximately constant [inset of fig.~\ref{fig2},a)]. The full width at half maximum (FWHM) of 27 $\mu$eV corresponds to the width of temperature broadened peaks at 90 mK \cite{Kouw01}. At larger gate voltage [inset of fig.~\ref{fig2},b)], the coupling to the leads becomes stronger and the dot enters an intermediate coupling regime ($h \Gamma \lesssim k_BT$). The peak width increases continuously as a function of gate voltage due to an increase of the level broadening. This shows directly the influence of the coupling strength on the peak width. Although the tunnelling rate of lead 1 increases more than the others [fig.~\ref{fig2},b)], the widths of all Coulomb peaks increase the same way. This means that the level broadening due to the coupling to one lead can be seen in the width of all the other conductances, as expected if all leads are coupled to the same state. Here we point out that this last result only holds on a qualitative level, since the dot is no longer in the single-level tunnelling regime. In the inset of fig.~\ref{fig3},d), the same analysis is carried out for the magnetic field dependence. The peak width does not change when increasing the magnetic field, although the tunnelling rates vary strongly, meaning that the dot is still in the weak coupling regime.

The implementation of spin-based quantum information processing \cite{Loss01,Loss02,Burkard01} with quantum dots requires more complex and coupled quantum systems. Two coupled quantum dots embedded in an Aharonov-Bohm interferometer are suggested for the detection of entanglement of spins states \cite{Loss02}. A three-terminal dot could be a step towards building an electronic entangler \cite{OliverSaraga}. For such devices, the quantitative control of individual coupling strengths will be a prerequisite for the desired functionality. We have demonstrated that a dot in the weak coupling regime (see fig.~\ref{fig2},a)), where tunnelling through a single level is expected to occur, can be tuned into a regime where the average coupling strengths to all leads are approximately the same. However, fluctuations of the shape of the wave function in the dot do not allow to control each coupling strength independently. Quantum rings, which have a more regular energy level spectrum rather than a chaotic one, even for a large number of electrons \cite{Fuhr01}, could be better candidates. In general, our method of measuring multi-terminal quantum dots provides a means to deduce the coupling strengths in more complex systems. It is also suitable for studying interference effects on the peak height statistics in chaotic quantum dots \cite{FoaTorres01,ChangFolk,PatelFolk} by working directly with the individual tunnelling rates rather than with the two-terminal conductance.

\acknowledgments
Financial support from the Swiss Science Foundation (Schweizerischer Nationalfonds) and from the EU Human Potential Program financed via the Bundesministerium f\"ur Bildung und Wissenschaft is gratefully acknowledged. We thank C. Lewenkopf and Y. Gefen for useful discussions.

\end{document}